\documentclass[aps,prl,reprint,showpacs,superscriptaddress]{revtex4-1}
\usepackage[english]{babel}
\usepackage{amsmath,amssymb,bbm,graphicx,color,comment,txfonts}
\usepackage[bookmarks=true,colorlinks,citecolor=blue,urlcolor=blue]{hyperref}
\usepackage{dsfont}

\newcommand{\br}{\mathbf{r}}

\renewcommand{\Im}{\mathop{\mathrm{Im}}}
\DeclareMathOperator{\tr}{tr}

\begin{document}

\date{\today}

\title{Vanishing density of states in weakly disordered Weyl semimetals}
\author{Michael Buchhold}
\affiliation{Department of Physics and Institute for Quantum Information and Matter, California Institute of Technology, Pasadena, CA 91125, USA}
\author{Sebastian Diehl}
\affiliation{Institute for Theoretical Physics, Universit\"at zu K\"oln,  D-509237 K\"oln, Germany}
\author{Alexander Altland}
\affiliation{Institute for Theoretical Physics, Universit\"at zu K\"oln,  D-509237 K\"oln, Germany}

\begin{abstract}
The Brillouin zone of the clean Weyl semimetal contains points at which the density
of states (DoS) vanishes. Previous work suggested that below a certain critical
concentration of impurities this feature is preserved including in the presence of
disorder. This result got criticized for its neglect of rare disorder fluctuations
which might bind quantum states and hence generate a finite DoS. We here show that in
spite of their existence these states are so fragile that their contribution
effectively vanishes when averaged over continuous disorder distributions. This means
that the integrity of the nodal points remains protected for weak disorder.
\end{abstract}

\maketitle

{\it Introduction.} -- The three dimensional Weyl semimetal is a paradigm of gapless
topological quantum matter. Its defining feature is the presence  of an even number
of topologically protected band touching points in the Brillouin zone. The linearly
dispersive behavior of  these Weyl points is attracting a lot of attention and has
put the system at the center of experimental~\cite{Xu613, Weng, Inoue2016, Xu2015z,
Belo2016, Xu2016z} and theoretical~\cite{Surface1, Surface2,
ChiralAnThe1, Fermiarc1, Fermiarc2, VishRev} studies of relativistic Fermi matter in
solid state physics contexts.

While individual Weyl nodes enjoy topological protection --- they can be moved in the
Brillouin zone but not individually destroyed --- the presence of singular band
touching points makes the system highly susceptible to perturbations away from the
clean, non-interacting limit. Specifically, the role played by static disorder has
been the subject of a partly controversial debate: renormalized
diagrammatic perturbation theory in $d=2+\epsilon$ dimensions \cite{Fradkin86b,
Fradkin86a,Goswami2011,GurarieWeyl, Syzranov2016b, Sbierski2017,Roy2014,Roy2016} evaluated at $\epsilon=1$
\cite{GurarieWeyl2, GurarieWeyl3}, and the mean field analysis of a nonlinear sigma
model approach \cite{Fradkin86b, Fradkin86a, Altland2015, Altland2016} predict the
existence of a critical disorder strength, $K_c$, below which disorder is irrelevant
and the system behaves effectively clean at large length scales. However, this
finding  is at variance with a complementary approach~\cite{Nandkishore:2014aa}
arguing that rare disorder configurations are capable of generating zero energy
states, leading to a finite density of states at the Dirac point. No matter how
small, this would rule out the DoS as an order parameter and compromise the
existence of a phase transition driven by disorder strength. Finally, the numerical analysis of the problem is met with various challenges. For example, lattice implementations categorically model even numbers of Weyl nodes, which in the presence of finite range correlated disorder are coupled. This makes it difficult to resolve the spectrum of individual nodes down to the lowest energies, and the current status~\cite{Sbierski2015,Sbierski2014,
Pixley2016a,Pixley2016b,Wilson2016,Pixley2015c, Shapourian,Soumya,Ostrovsky2017,Kobayashi2014,Shang2016} does not appear to be fully conclusive.

In this paper we analytically demonstrate that the nodal DoS in the weakly disordered
Weyl semimetal remains  vanishing, including if rare fluctuations are taken into
account. At first sight this may sound counterintuitive. Rare fluctuations include
configurations in which the chemical potential of the system is effectively lowered
(or raised) over sizable regions in space. One might expect this to shift
the  Weyl cone away from the state-less nodal point with vanishing DoS and effectively accumulate finite
spectral weight at zero energy. In the following, we will analyze three different model setups to demonstrate that this is not what is happening. Specifically, we will a) analytically compute the DoS of a box potential mimicking a rare event fluctuations, b) apply the large fluctuation stationary phase methods to compute the DoS of rare Gaussian fluctuations, and c) consider a $T$-matrix approach to the DoS of multiple point like impurities. While these setups lead to distinct spectral density profiles and call for different computational schemes they all have in common that the spectral density at the nodal points remains vanishing.

\emph{Box potential ---} To start with consider a spherical potential of
radius $b$ and uniform depth $\lambda$ as a cartoon of a rare configuration. Unlike
in a Schr\"odinger problem, such wells bind states only at zero energy and only for
depths, $\lambda_c \equiv m \pi/b$, where $m$ is integer \cite{Pieper1969, Greiner}.
Away from these singular configurations, the fragile bound states turn into
scattering states whose interplay with the continuum of extended states is key to the
understanding of the DoS: according to the Levinson theorem (a close cousin of the
Friedel sum rule) \cite{Friedel,PRB}, the number of bound states in a
scattering setup equals the  phase shift difference, $\frac{2}{\pi}\int d\omega \,
\partial_\omega\delta(\omega)$ accumulated by all scattering states. (In the Weyl
problem, the integral extends over positive and negative energies, up to a cutoff
beyond which the potential is no longer effectively seen.) This implies that
 no additional DoS is generated by a  potential and  that 
accumulations of density of states in one energy region, i.e. by a bound state or
resonance, are screened by a diminished scattering state $\delta\nu(\omega)=\frac{2}{\pi}\partial_\omega\delta(\omega)$ elsewhere.

\begin{figure}
	\includegraphics[width=\linewidth]{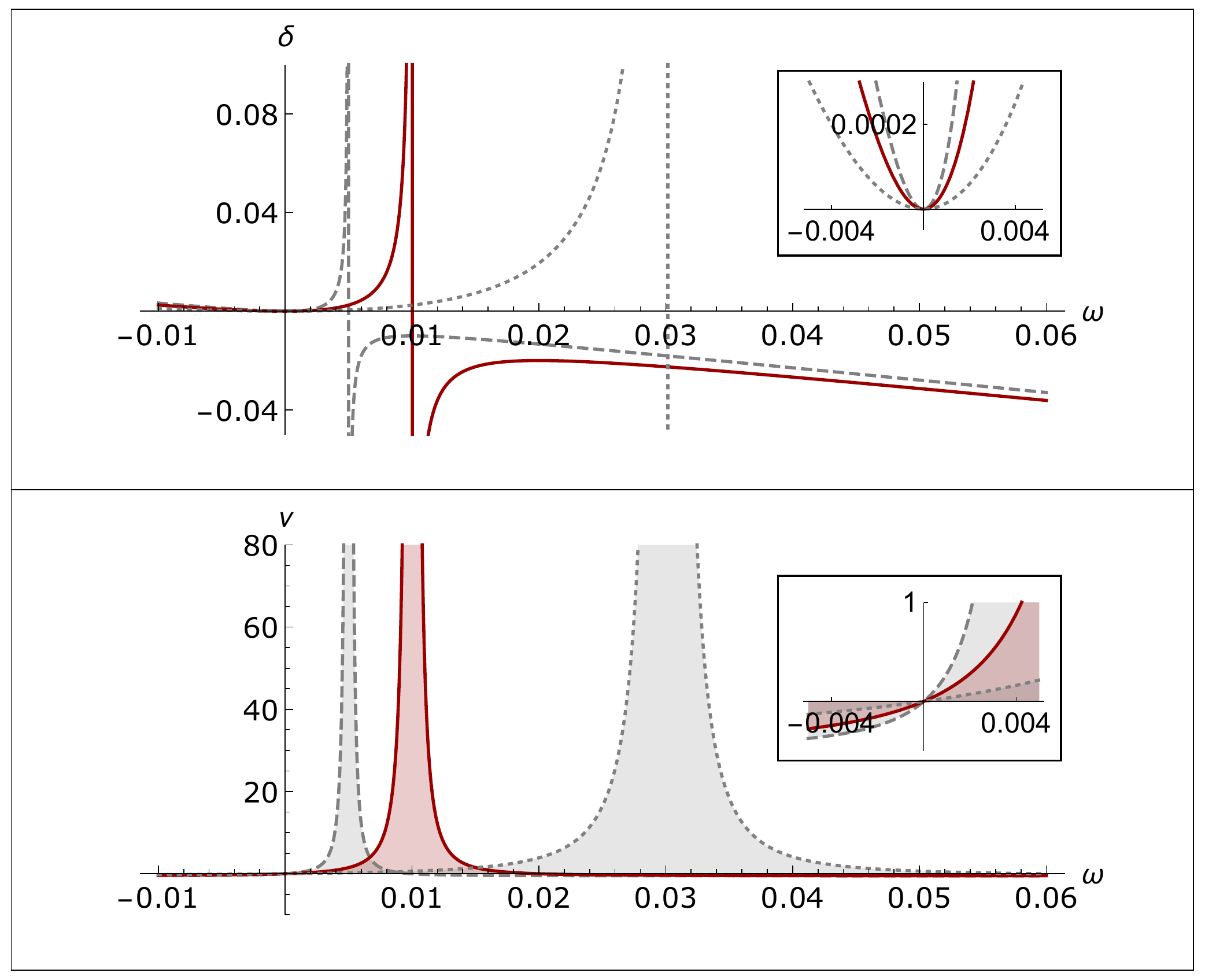}
	\caption{Top: phase shift $\delta(\omega)$ for three configurations, $\lambda=\pi/b+\Delta$, $\Delta=0.01$ (dashed), $0.02$ (solid), $0.06$ (dotted). With decreasing $\Delta$, a resonance of width $\sim \Delta$ moves closer to the origin. At the origin, $\delta(0)=\delta'(0)=0$ touches zero (inset). Bottom: the corresponding  shift in the DoS $\delta\nu\equiv \nu_\lambda-\nu_0$ compared to the clean Weyl problem assumes the form of narrow peaks. Their positive spectral weight is screened by a negative background in such a way that $\nu(0)=0$, always (inset). This includes the  limit, $\Delta\to 0$, in which the singular derivative of the phase shift reflects the presence of a bound state.}
	\label{Fig1}
\end{figure}

The explicit diagonalization of the potential well Hamiltonian \cite{Friedel,
Pieper1969} shows that for $\lambda=\lambda_c$ the phase drop $\pi \to 0$
compensating for the zero energy bound state is concentrated in the infinitesimal
neighborhood of zero energy. In effect, this means that \emph{no} spectral weight
accumulates there. Tuning away from the singular value, $\lambda_c$, the bound state
becomes a  finite energy scattering state. However, the ensuing narrow peak,
$\delta\nu(\omega)$, is screened by a negative background (see Fig.~\ref{Fig1}), in
quantitative agreement with the sum rule. The explicit calculation~\cite{PRB}
shows that  the balance is such that $\nu(0)=0$ for all parameter values:  a
potential well does generate spectral density, but never at zero energy. This is the
reason for the absence of a rare event contribution to the DoS in the Weyl problem.
One might object that the argument makes reference to a potential well of specific
shape (a box) and does not account for potential side effects due to
\emph{correlations} between neighboring potential inhomogeneities. In the following,
we will show that the vanishing of the DoS is robust and survives these
generalizations.

{\it Random potential wells ---} We follow Ref.~\cite{Nandkishore:2014aa}, and
analyze the rare event DoS in a finite range Gaussian correlated potential via a
standard~\cite{LifshitzRev1, LifshitzRev2, Lutt,Rossum1994,Langer66,Yaida2016} large
deviation analysis. Referring to Ref.~\cite{PRB} for details, we limit
ourselves to a sketch of the  construction  and explain where we deviate  from
Ref.~\cite{Nandkishore:2014aa}. The starting point is a representation of the DoS as
a Gaussian functional integral
\begin{align}
  \label{DoSRepr}
    \nu(\omega) = -\frac{1}{2\pi L^3}\mathrm{Im}\int \mathcal{D}[\bar\psi,\psi]\,\int (\bar\phi \phi-\bar\chi\chi)\left\langle e^{i S[\psi]}\right \rangle,
\end{align}
where $S[\psi]\equiv \int \bar\psi (\omega^+-\hat H)\psi$, $\psi=(\phi,\chi)^T$ is a field comprising
commuting and anticommuting components, $\phi$ and $\chi$, respectively, and $\hat H = -i v_0
\sigma_i\partial_i+V_x$ contains the Weyl-Hamiltonian with velocity $v_0$ and a Gaussian potential, $\langle V_x
V_{x'}\rangle =W^2\exp(-|x-x'|/\xi)$ of strength $W$ and correlation length, $\xi$,   mimicking the formation of finite range potential wells. 

The auxiliary integration over the anti-commuting $\chi$ is required to avoid the unwanted appearance of determinants $\det(\omega^+-\hat H)$ after the integration~\cite{EfetovBook}. In tis case, the average over disorder simply generates an effective action
\begin{align}
    \label{eq:EffectiveAction}
     S_\mathrm{eff}[\psi]=\int_x \,\bar \psi_x\Big(\omega^++iv_0\sigma_i\partial_i+iW^2\int_{y}e^{-\xi^{-1}|x-y|}\bar\psi_{y}\psi_{y}\Big)\psi_x,
 \end{align}
quartic in integration variables. We are now at the crossroad where the two principal approaches to computing the DoS part: where Refs.~\cite{Fradkin86a, GurarieWeyl} apply renormalized perturbation theory to the analysis of the quartic nonlinearity, the starting point of Ref.~\cite{Nandkishore:2014aa} is the observation~\cite{Lutt,Rossum1994,Langer66} that the physics of rare events is contained in inhomogeneous `instanton' solutions to the variational equations of the action~\eqref{eq:EffectiveAction}.

Following the second approach, we consider nontrivial solutions, $\psi_I$, of the
variational equation, $\delta_\psi S =0$ identified in
Ref.~\cite{Nandkishore:2014aa}. Referring to the original reference
and~\cite{PRB} for details we note that a class of instanton solutions with
regular behavior at the origin and power law decay at large scales, $r>\xi$ can be
identified. Reference~\cite{Nandkishore:2014aa} reasoned that, to exponential
accuracy, the value of the DoS close at zero energy should be determined by the instanton action, $\nu\sim \exp(-S[\psi_I])\sim
\exp(-v_0^2/(W\xi)^2)$. We here take the additional step to include quadratic
\emph{fluctuations}  around the instanton saddle point. That fluctuations may be less
innocent than in conventional large deviation phenomena is indicated by the
fragility of the bound states discussed above: one may suspect that in spite of the
finite probability to find bound states, their singular sensitivity to parameter
variations leads to a vanishing measure. Within an integral approach, this would show
in a vanishing fluctuation contribution around extremal configurations.

The extremal solutions $\psi_I$ break seven continuous symmetries: translational invariance in three directions, three independent rotations in the complex two-component Weyl space, and one supersymmetry. While the first six are harmless and can be treated according to standard procedures in instanton calculus \cite{Brezin1977}, the supersymmetry breaking deserves more attention: consider a commuting solution of the stationary equations, $\psi_I=(\phi_I,0)$, where $\phi_I$ is a two-component complex Weyl spinor. If we act on this configuration with a rotation in super-space, $\psi_I\to W\psi_I$, where $W=\exp\left( \begin{smallmatrix}
    0&\eta\mathds{1}\crcr \bar \eta\mathds{1} &0
\end{smallmatrix} \right) $,  $\mathds{1}$ is the unit-matrix in Weyl space, and $\eta,\bar\eta$ are Grassmann variables, a new solution is generated. (The matrix $W$ commutes with the Hamiltonian, which is the supersymmetry.) This means that $\eta,\bar \eta$ are Grassmann-zero modes, and that the generalization $\eta\to \eta(x)$ to fields with slow coordinate dependence generates soft fluctuation modes. The next step in the analysis is to expand the action around $\psi_I$ to at least quadratic order in fluctuation modes. Expanding a general Grassmann fluctuation as $\eta(x)=\sum_a F_a(x)\eta_a$ in a set of suitably defined functions, $F_a$, this leads to an expression of the symbolic structure $S[\eta]=\sum_a \bar \eta_a X_{ab}\eta_b$ governed by an effective fluctuations operator. After integration over $\eta_a$ the integral picks up a factor $\det(X)$ which, unlike with commuting fluctuation variables, appears in the numerator of the fluctuation prefactor.
The operator $X$ contains at least one zero eigenvalue, whose eigenmode is the
constant fluctuations $\eta_0=\mathrm{const}$. In the computation of the density of
states this factor is canceled after the expansion of the pre-exponential term $\bar
\psi\psi$ in the Grassmann variables $\{\eta_a\}$ and integration. This $0/0$
cancellation is of general nature and safeguards the correct normalization of
observables~\cite{EfetovBook}. However, should a fluctuation operators contain $n>1$ one zero eigenvalues, the integration leads to a factor $0^n/0$ and a vanishing result.

The setting as described so far is of general nature and in the same way applies to,
e.g., a disordered Schr\"odinger operator. What makes the Weyl problem special is
that at $\omega=0$ it indeed possesses an extensive (diverging in the limit of
infinite volume) number of fluctuation modes $\eta_a$ of vanishing action, and the
above structure implies a vanishing of the rare event contribution to the DoS. The
existence of these modes simply follows from the fact that  the Weyl operator is of first order in
derivatives. The condition of vanishing fluctuation action, thus assumes the form $\hat O \eta=0$, where $\hat O$ is a first order partial differential
operator.  Unlike with equations governed by second order elliptic operators (as in a Schr\"odinger problem) these first order differential equations possess an extensive number of solutions which can be found, e.g., via the method of characteristics~\cite{PRB}. In the  asymptotic limit of a perfectly linear
Weyl operator and a single instanton defined in a box of unbounded extension, $L\to
\infty$ we indeed obtain a diverging number of zero fluctuation zero modes, and hence
a vanishing DoS. The large spatial support of these fluctuation modes  reflects the absence of compact exponentially bound eigenstates, as exemplified above for the box potential. 

A realistic Weyl operator is not perfectly linear but contains higher order
derivatives due to, e.g., an underlying lattice dispersion. (Higher order derivatives
in the fluctuation action are also induced via the coupling between the zero modes
$\eta$ to other fluctuation modes.) We also have to account for  multi-instanton
solutions containing the superposition of inhomogeneities $\psi_I$ centered around
different coordinates.  These generalizations affect the
above analysis via the appearance of an effective length scale, $L\sim \mathrm{min}(
L_I, \Gamma^{-1})$, loosely to be identified with the instanton separation, $L_I$ or
the inverse of momentum scale, $\Gamma$, above which the spectrum ceases to be
linear. The introduction of this scale lifts the zero degeneracy of the fluctuation
modes to a spectrum $\epsilon_n \sim L^{-1}F(n)$, where $F(n)$ is a polynomial factor
containing some `quantum numbers' $\{n_i\}$ describing the angular momentum and
radial quantization of the fluctuation modes. The fluctuation determinant contains
the product $\prod_n \epsilon_n$, and a quick estimate shows that the product up to
values $\epsilon_n\sim 1$ leads to a factor $\sim \exp(-cL)$, where
$c=\mathcal{O}(1)$. This factor sets an upper bound for the DoS, as obtained by the present approach. For example, if $L\sim \exp(S_I)$ is identified with the expected separation between instanton configurations, we obtain a bound $\sim \exp(-c \exp(v_0^2/W^2\xi^2))$ double exponential in the disorder concentration. To summarize, the inclusion of fluctuations leads to a drastic suppression of the DoS compared to the fluctuationless stationary phase analysis. However, in view of approximative assumptions --- linearity of the spectrum, neglect of correlations --- required for the computation of fluctuation determinants, the DoS cannot be determined with ultimate precision.

\emph{Multiple impurities ---} We now discuss a model that does account for potential
correlations and does tolerate (even require) nonlinear dispersion. The price to be
payed for the enhanced degree of generality is that the potential landscape is
modeled in a simplistic fashion: following Ref.~\cite{Ziegler2017}, we consider a system of
$N$ point-like impurities, $V(\br)=\sum_i U_i \delta(\br-\br_i)$ at positions
$\br_i$ and strength $U_i$. Here, the $\delta$-potentials are a cartoon of
fluctuations whose range $\xi>b^{-1}$ is larger than the momentum separation between
different Weyl nodes so that trans-nodal scattering is excluded, and at the same time
$\xi<M^{-1}$ smaller than an effective large momentum cutoff of individual nodes.
Within this parametric window, the spatial profile of individual minima is not
resolved and a point-like model  justified. At the same time, the model is simple
enough to be exactly solvable by $T$-matrix techniques.

\begin{figure}
    \includegraphics[width=\linewidth]{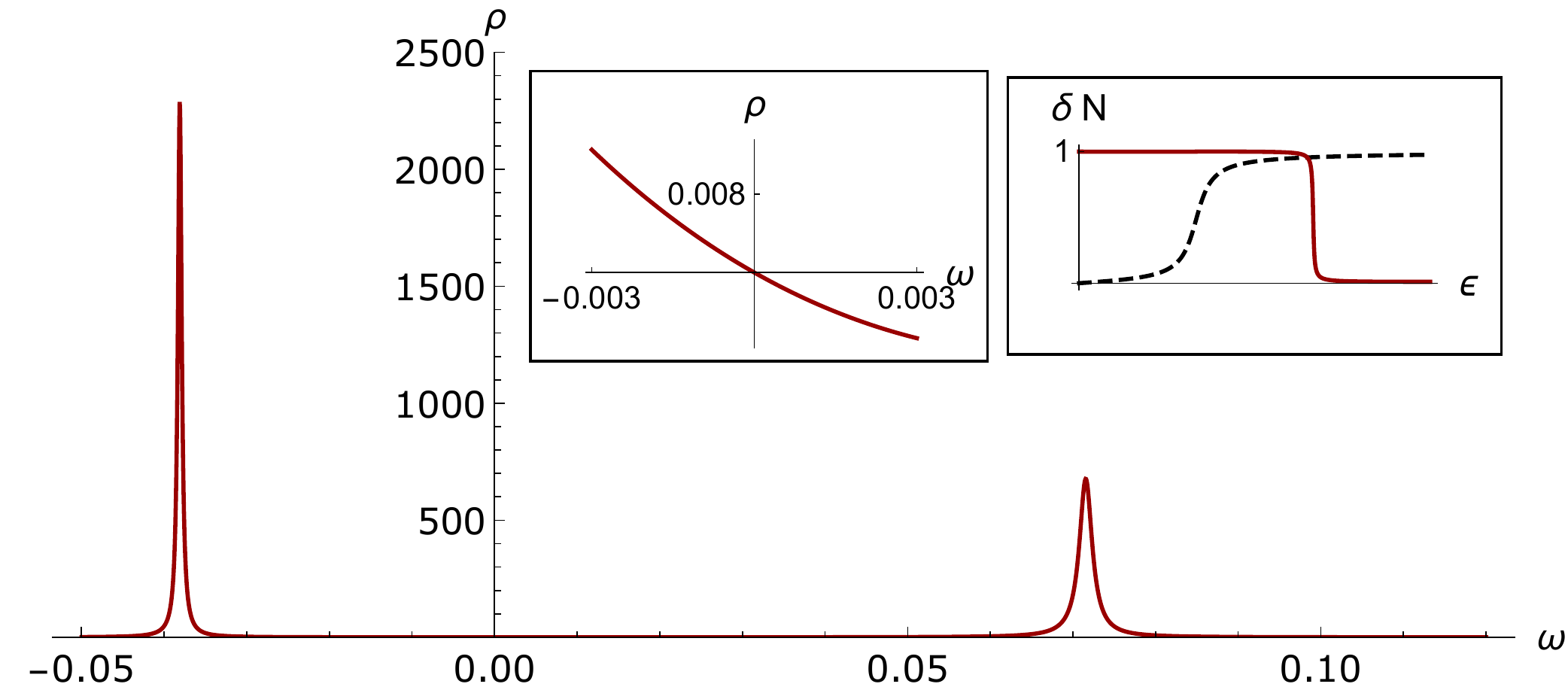}
    \caption{Exemplary DoS of a system of two impurities ($r=15/M$, $U_1=-75/M^2$, $U_2=150/M^2$ in the units discussed in the text). Left inset: blowup of the DoS near zero energy. Right inset: dashed, integrated spectral weight around the resonance centers, solid, the same for a resonance at zero energy.}
    \label{Fig2}
\end{figure}

An expansion of the Green function $\hat G=(\omega^+-\hat H_0-\hat V)^{-1}$ shows
that the impurity contribution to the DoS is given by
$\delta\nu(\omega)=-\frac{1}{\pi}\Im\tr \left(\hat G_0 (1-\hat V\hat G_0)^{-1}\hat
V\hat G_0\right)$, where $\hat G_0$ is the clean Green function. Introducing a
diagonal matrix $U=\mathrm{diag}(U_1,\dots,U_N)$ and the projection operator $\hat
P=\sum_i |\br_i\rangle \langle \br_i|$ onto impurity coordinates, this can be
represented as $\delta\nu(\omega)=-\frac{1}{\pi}\partial_\omega \Im F(\omega)$, where
the `free energy' $F(\omega)=\ln \det (\hat U^{-1}-\hat G_0)$ contains the projected
Green function $\hat G_0=\hat P G_0\hat P$. In this way, the computation of the DoS
is reduced to that of the determinant of an $N\times N$ matrix. The projected Green
function matrix elements, $G(\br_i-\br_j)$ featuring in this matrix contain the
singular diagonals, $G(0)=\int
\frac{d^3k}{(2\pi)^3}\frac{\omega^+-k_i\sigma_i}{\omega^{+2}-k^2}$ (we set $\nu_0\equiv1$ here). While this
expression can be regularized in different ways, the introduction of nonlinearity in
the band dispersion, $\epsilon(k)=v |k|+\mathcal{O}(k^2/M^2)$, where $M$ is an effective large momentum cutoff, may be the most natural
for a system defined on a lattice. Either way, one obtains $G(0)=-\frac{\omega^+
(M+i\omega)}{4\pi} +\mathcal{O}(\omega^2)$. The off-diagonal elements are
non-singular, and for small $\omega$ assume the form $G(\br)=-\frac{ir_i\sigma_i+r^2
\omega}{4\pi r^3}+\mathcal{O}(\omega^2)$. 

It is instructive to explore the ensuing DoS profile for a system of just two
impurities at distance $r$. In this case,
$F(\omega)=-2\ln\left((U_1^{-1}+M\omega^+)(U_2^{-1}+M\omega^+)-r^{-4}-r^{-2}\omega^{+2}\right)$
and the DoS near zero is obtained as the asymptotically linear function
$\delta\nu(\omega)=\frac{r^4 (U_1 +U_2)\omega}{\pi^2 (U_1
U_2-r^4)}+\mathcal{O}(\omega)^2$. Away from zero energy the DoS shows resonances at
energies $\omega_0=-\frac{1}{2MU_1 U_2}(U_1+U_2 \pm
\left((U_1-U_2)^2+4r^{-4}U_1^2U_2^2\right)^{1/2})$, which are the resonant energies
$1/MU_i$  of the isolated impurities, shifted by an $r$-dependent
hybridization energy. The peak values at resonance diverge in $M$ and the
width shrinks in the inverse of the same parameter. Each resonance carries unit spectral weight, as indicated in the right inset of Fig.~\ref{Fig2}, where the integrated spectral density $N_\epsilon\equiv \int_{\omega_0-\epsilon}^{\omega_0+\epsilon}d \omega
\delta\nu(\omega)=-\frac{1}{\pi}\mathrm{Im}F(\omega)\big|^{\omega_0+\epsilon}_{\omega_0-\epsilon}$ is shown as a dashed curve. 

Setting $\omega_0\to 0$, we identify
the configurations $U_1U_2=r^4$ for which the impurity hybridization pushes the
resonance centers to zero. In this limit,
the slope of the DoS $\partial_\omega \delta\nu(\omega)$ diverges and
$\lim_{\omega\to 0}\delta\nu(\omega)$ is no longer defined. However,  the integrated spectral weight $N_\epsilon$ is still well defined and inspection of the logarithm shows that $N_\epsilon \rightarrow 0$ for energies $\epsilon\gtrsim
\epsilon_0\equiv (U_1^{-1}+U_2^{-1})/M$. This is shown in the right inset Fig.~\ref{Fig2}, where the position of the final kink is set by $\epsilon_0$. The structure indicates that  the DoS  carried by the zero energy peak is `screened' by an equally strong counterweight in its immediate vicinity, $\epsilon_0\sim M^{-1}$. In the asymptotic limit of a fully linear spectrum, $M\to \infty$, the zero energy resonance does not carry spectral weight at all, $\delta N_\epsilon=0$, and for finite $M$, $\delta N_{\epsilon}$ shows sign-fluctuating singular behavior at energies $\omega\lesssim \epsilon_0(M)$. However, regardless of the value of $M$, the above singular profile will never be realized in any specific sample of impurity potentials. The reason is that upon approaching zero energy the resonant peaks not only become larger they also become \textit{narrower}. A careful statistical analysis of the problem~\cite{PRB} shows that the limit of a singular zero energy resonance is an event of measure zero in the sense of probability theory, and is strictly non-observable.

{\it Summary and Outlook.} -- Summarizing, we have analyzed three different models of
disorder in Weyl semimetals, all preserving the integrity of the nodal point.
Individual of these models  emphasized different facets of the problem. Specifically,
the box potential model was simple enough to be amenable to exact analytic solution
by scattering methods. The model of Gaussian distributed disorder could no longer be
solved rigorously, in exchange for a more realistic modeling of a smooth disorder
landscape. Finally, the multi-impurity model described individual impurities in
simplistic ways, but added the effects of impurity correlations and spectral
curvature to the analysis. The fact that three different models and different
analytic approaches lead to identical conclusions indicates that the protection of
the nodal structure is a general result. 

We have seen that in all three models the DoS away from zero energy is carried by a peculiar set of resonances. While nothing prevents these resonances from approaching zero, they become narrower (and hence more difficult to observe) in the process. For any finite separation from zero, the nodal DoS  remains continuous and vanishing. Only in the limit, the competition of diverging resonance height and vanishing width leads to a singularity. However, this limit has zero statistical measure and is not realized in any specific sample (much like a mathematical zero will not be drawn in any random sampling of real numbers.) This is how the seeming contradiction between zero nodal DoS and resonant DoS elsewhere gets resolved. The detailed analysis of the ensuing statistical DoS distribution is a subject of Ref.~\cite{PRB}.


Finally, the principal result of preserved nodal points is a result of
conceptual significance. It suggests that the mean field result of a threshold
concentration separating a weak and a strong disorder phase survives the presence of
rare events, and that there is a genuine phase transition with $\nu(0)$ as its order
parameter.

{\it Acknowledgements.} --- We thank P. W. Brouwer, V. Gurarie, R. Nandkishore, L. Radzihovsky, G. Refael, B. Sbierski, J. H.
Wilson, K. Ziegler, and M. Zirnbauer for discussions. Work supported by the German Research
Foundation (DFG) through the Institutional Strategy of the University of Cologne within the German Excellence
Initiative (ZUK 81) and CRC/TR 183 -- Entangled states of matter (project A02). M.~B. acknowledges support from the Alexander von Humboldt foundation.
\bibliography{Diss}

\end{document}